\documentclass[aps,notitlepage,superscriptaddress,twocolumn,a4paper,floatfix,nofootinbib]{revtex4-1}

\usepackage{graphicx}

\newcommand{\co}{c_{\rm o}}
\newcommand{\eg}{{\it e.g.}, }
\newcommand{\erup}{\epsilon_{\rm rup}}
\newcommand{\etal}{{\it et al.\ }}
\newcommand{\gexp}{g_{\rm exp}}
\newcommand{\ie}{{\it i.e.}, }
\newcommand{\kB}{k_{\rm B}}
\newcommand{\pd}{\partial}
\newcommand{\po}{p_{\rm o}}
\newcommand{\Qc}{Q_{\rm c}}

\newcommand{\tM}{\tilde{M}}
\newcommand{\vw}{v_{\rm w}}

\begin{document}

\title{Law of corresponding states for osmotic swelling of vesicles}

\author{Primo\v{z} Peterlin}
\altaffiliation[Present address: ]{Institute of Oncology, Zalo\v{s}ka 2, 1000 Ljubljana, Slovenia}
\affiliation{Faculty of Medicine, Institute of Biophysics, University of Ljubljana,
Lipi\v{c}eva 2, 1000 Ljubljana, Slovenia}

\author{Vesna Arrigler}
\affiliation{Faculty of Medicine, Institute of Biophysics, University of Ljubljana,
Lipi\v{c}eva 2, 1000 Ljubljana, Slovenia}

\author{Emir Haleva}
\affiliation{Raymond \& Beverly Sackler School of Chemistry, Tel Aviv University,
Tel Aviv 69978, Israel}

\author{Haim Diamant}
\email{hdiamant@tau.ac.il}
\affiliation{Raymond \& Beverly Sackler School of Chemistry, Tel Aviv University,
Tel Aviv 69978, Israel}

\begin{abstract}
  As solute molecules permeate into a vesicle due to a concentration
  difference across its membrane, the vesicle swells through osmosis.
  The swelling can be divided into two stages: (a) an ``ironing''
  stage, where the volume-to-area ratio of the vesicle increases
  without a significant change in its area; (b) a stretching stage,
  where the vesicle grows while remaining essentially spherical, until
  it ruptures. We show that the crossover between these two stages can
  be represented as a broadened continuous phase transition.
  Consequently, the swelling curves for different vesicles and
  different permeating solutes can be rescaled into a single,
  theoretically predicted, universal curve.  Such a data collapse is
  demonstrated for giant unilamellar POPC vesicles, osmotically
  swollen due to the permeation of urea, glycerol, or ethylene glycol.
  We thereby gain a sensitive measurement of the solutes' membrane
  permeability coefficients, finding a concentration-independent
  coefficient for urea, while those of glycerol and ethylene glycol
  are found to increase with solute concentration.  In addition, we
  use the width of the transition, as extracted from the data
  collapse, to infer the number of independent bending modes that
  affect the thermodynamics of the vesicle in the transition region.
\end{abstract}

\maketitle

\section{Introduction}
\label{sec_intro}

Membrane vesicles are made of a closed bilayer of amphiphilic
molecules in aqueous solution, having length scales of $0.1$--$100$
$\mu$m.  Used as simplified models of biological membranes, they have
been one of the most extensively studied systems in soft-matter
physics \cite{Safran,Seifert1997}. Vesicles usually enclose both
solvent (water) and solute molecules. Such vesicular capsules are
ubiquitous in cell functions \cite{bio} and used as microreactors and
delivery vehicles in various biomedical and cosmetic applications
\cite{Lasic}.

The hydrophobic core of the bilayer membrane poses a kinetic barrier
to the permeation of water and water-soluble molecules into and out of
the vesicle. Consequently, the permeability coefficients of various
solutes across various membranes span a very wide range of values
\cite{Paula1996}. For example, the permeability coefficients of water,
urea, and potassium ions through the membranes used in the current
work are of the order, respectively, of $10^2$, $10^{-2}$, and
$10^{-8}$ $\mu$m/s \cite{Paula1996}. This implies that the time scales
for permeation of water and the solutes examined here are well
separated. Thus, the vesicles can safely be assumed to reside in a
semi-permeable regime, in which their volume quickly (essentially
immediately) re-adjusts through osmosis to a change in the number of
enclosed solute molecules. As more solute molecules enter, therefore,
the vesicle progressively swells\,---\,first approaching a spherical
shape (the ``ironing'' stage), subsequently inflating as a sphere (the
stretching stage), and eventually rupturing (osmotic lysis)
\cite{Mui1993}.

In an earlier theoretical work we argued that a vesicle in such a
semi-permeable regime should reach the end of the ironing stage (\ie
the maximum volume-to-area ratio) critically, through a continuous
transition \cite{PRL08}. That theory was restricted to unstretchable
membranes. Here we extend the theory to the experimentally relevant
case of stretchable membranes, finding a slightly modified but similar
form of criticality. We then confirm the existence of the critical
scaling behavior in experiments and utilize it to obtain a reliable
measurement of permeability coefficients.

A crossover between two stages of vesicle swelling is well known in
micropipette-aspiration experiments \cite{Evans1990}. The stages are
distinguished by a markedly different dependence of surface tension on
surface strain\,---\,in the first stage the dependence is exponential,
while in the second it is linear
\cite{Evans1990,Helfrich1984,Milner1987}. The essential difference
between this scenario and the one addressed here lies in the different
control parameters. In the former case the swelling is controlled by a
hydrostatic pressure difference, whereas in the latter it is
controlled by the number of encapsulated solute molecules. As a
result, the two crossovers are not equivalent. As will be presented
below, in fact, the monitored swelling of our vesicles, with its two
distinct stages, occurs within the linear regime\,---\,\ie the tension
depends linearly on strain throughout the observed transition.

We begin by describing the experimental setup in Sec.\ \ref{sec_exp}
and the direct experimental results in Sec.\ \ref{sec_results}. The
analysis of the experimental data requires a revised theory, which is
presented in Sec. \ref{sec_theory}. In Sec.\ \ref{sec_analysis} we
apply the theory to the experimental results to demonstrate data
collapse onto a master curve\,---\,\ie the law of corresponding
states\,---\,and extract additional information, such as the
permeability coefficients and the number of surface bending modes
contributing to vesicle thermodynamics. Finally, we discuss the
various results and their significance in Sec.\ \ref{sec_discuss}.

\section{Experimental setup}
\label{sec_exp}


D-(+)-glucose, D-(+)-sucrose, glycerol, urea, and ethylene glycol were
purchased from Fluka (Buchs, Switzerland).  Methanol and chloroform
were purchased from Kemika (Zagreb, Croatia).
1-palmitoyl-2-oleoyl-\textit{sn}-glycero-3-phosphocholine (POPC) was
purchased from Avanti Polar Lipids (Alabaster, USA).  All the
solutions were prepared in double-distilled sterile water.


A suspension of POPC giant unilamellar vesicles (GUVs) in 0.1 or
0.2~mol/L 1:1 sucrose/glucose solution was prepared using an
electroformation method, described in Ref.\ \citenum{Angelova:1986}
with some modifications \cite{Heinrich:1996,CSB08}.  Lipids were dissolved
in a mixture of chloroform/methanol (2:1, v/v) to a concentration of
1~mg/mL.  A volume of 25~$\mu$L of the lipid solution was spread onto
a pair of Pt electrodes and dried under reduced pressure (water
aspirator; $\approx 60$~mmHg) for 2~hours.  The electrodes were then
placed into an electroformation chamber, which was filled with 0.1 or
0.2~mol/L sucrose.  AC current (8~V, 10~Hz) was applied, and the
voltage and frequency were reduced in steps to the final values of 1~V
and 1~Hz \cite{CSB08}.  Subsequently, the chamber was first drained
into a beaker and then flushed with an equal volume of isomolar
glucose solution, thus resulting in a suspension of GUVs containing
entrapped sucrose in a 1:1 sucrose/glucose solution, which increases
the contrast in a phase contrast setup and facilitates vesicle
manipulation \cite{Dimova:2006}.  This procedure yields mostly
spherical unilamellar vesicles, with diameters of up to 100~$\mu$m.


An inverted optical microscope (Nikon Diaphot 200, objective 20/0.40
Ph2 DL) with micro-manipulating equipment (Narishige MMN-1/MMO-202)
and a cooled CCD camera (Hamamatsu ORCA-ER; C4742-95-12ERG), connected
{\it via} an IEEE-1394 interface to a PC running Hamamatsu Wasabi
software, was used to obtain phase contrast micrographs.  In the
streaming mode, the camera provides $1344\times 1024$ 12-bit grayscale
images at a rate of 8.9~images/s.

In the experiment, an individual spherical GUV is selected, fully
aspirated into a glass micropipette whose inner diameter exceeds the
vesicle's diameter, and transferred from a solution containing solutes
of very low membrane permeability (1:1 glucose/sucrose) into an
iso-osmolar solution of a more permeable solute (glycerol, urea, or
ethylene glycol), where the content of the micropipette is released,
and the micropipette is subsequently removed.  Vesicle response is
recorded using a CCD camera mounted on the microscope. The radius of
the vesicle's cross-section was determined from the recorded series of
images 
using a least-squares procedure \cite{Peterlin:2009a}.

\section{Experimental results}
\label{sec_results}

\begin{figure}[tbh]
\centerline{\resizebox{0.4\textwidth}{!}{\includegraphics{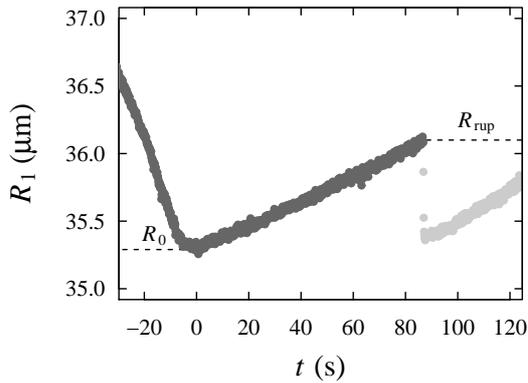}}}
  \caption{Typical swelling curve, showing the change in the observed
    (larger) radius of an oblate spheroidal POPC vesicle as a function
    of time due to inward permeation of urea. The decreasing part of
    the curve corresponds to the ironing stage, and the increasing
    one to the stretching stage. The minimum of the curve is defined
    as $(t=0,R_1=R_0)$. The present analysis concerns the swelling up
    to the point of first rupture ($R_1=R_{\rm rup}$).}
\label{fig_swellingcurve}
\end{figure}

Once the vesicle is released and gets in contact with the target
solution, a transient increase of its cross-sectional radius, $R_1$,
is observed. We attribute it to a small hypotonicity of the outer
solution, causing the vesicle to slightly deflate and change its shape
from a sphere into an oblate spheroid
\cite{Dobereiner:1997,Linke2005}. Gravity should lead to a small
deviation of the shape from a perfect oblate spheroid
\cite{Dobereiner:1997}, which we neglect here. A more important effect
of gravity is the breaking of the problem's rotational symmetry, such
that the observed lateral radius, $R_1$, is always the spheroid's
larger radius. On a longer time scale than the transient deflation,
the vesicle inflates due to solute permeation and the accompanying
osmosis (\ie water influx). This causes the vesicle to become more
spherical, making the observed $R_1$ decrease. (See Fig.\ 
\ref{fig_swellingcurve}.) This is the ironing stage, where the
swelling increases the volume-to-area ratio. After a certain time,
$R_1$ reaches a minimum and starts increasing. (See again Fig.\ 
\ref{fig_swellingcurve}.) This marks the crossover to the stretching
stage, where the vesicle continues to swell essentially as an
inflating sphere.  When the membrane reaches a critical strain, it
ruptures, and the vesicle bursts. Subsequently, the membrane is
resealed and another cycle of swelling commences. The repeated
burst--swelling cycles were analyzed elsewhere \cite{CSB08}. In the
present work we focus on the swelling that precedes the first burst.

The ironing and stretching stages are clearly distinguished in Fig.\
\ref{fig_swellingcurve} as the decreasing and increasing parts of the
swelling curve, respectively. To have a systematic definition of the
time axis for all vesicles we define $t=0$ at the minimum of the
curve. We associate the radius at that minimum, $R_0$, with the
relaxed area of the vesicle, $A_0\equiv 4\pi R_0^2$. (We return to
examine this assumption in Sec.\ \ref{sec_discuss}.)  From each such
curve we also extract the radius at rupture, $R_{\rm rup}$. This
yields a direct measurement of the membrane's rupture strain,
\begin{equation}
   \erup = (R_{\rm rup}/R_0)^2 - 1.
\label{strain}
\end{equation}

\begin{figure}[tbh]
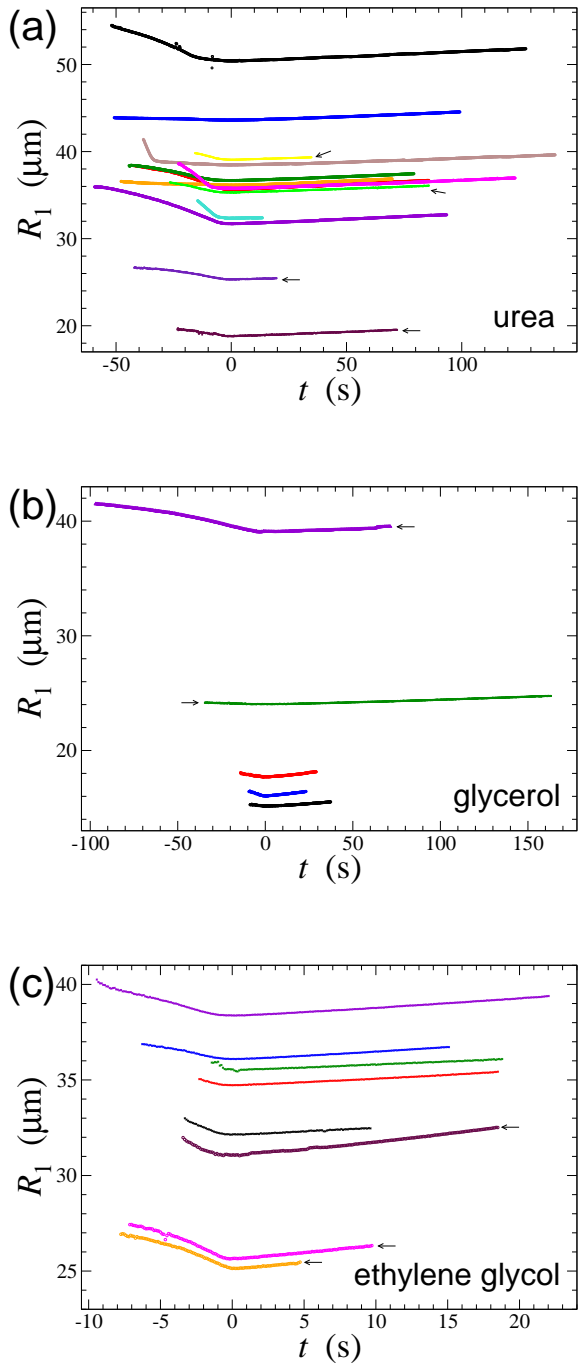

\vspace{0.5cm}
\centerline{\resizebox{0.42\textwidth}{!}{\includegraphics{fig2a.eps}}}
\vspace{1cm}
\centerline{\resizebox{0.42\textwidth}{!}{\includegraphics{fig2b.eps}}}
\vspace{1cm}
\centerline{\resizebox{0.42\textwidth}{!}{\includegraphics{fig2c.eps}}}
\caption{Swelling curves for POPC vesicles under the permeation of (a)
  urea, (b) glycerol, and (c) ethylene glycol. Outer solute
  concentrations are 0.1~M, 0.2~M (left-pointing arrows), and 0.11~M
  (right-pointing arrow in (b)).}
\label{fig_swellingcurves}
\end{figure}

\begin{figure}[tbh]
\vspace{0.5cm}
\centerline{\resizebox{0.4\textwidth}{!}{\includegraphics{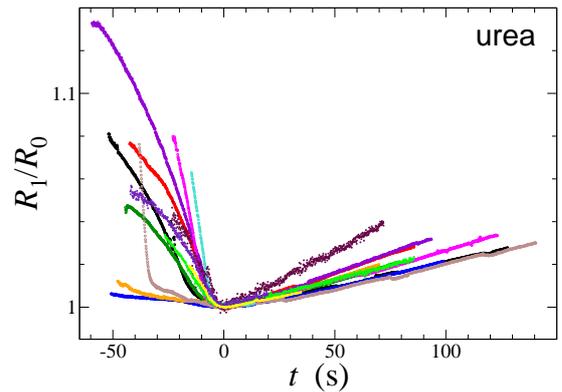}}}
\caption{The swelling curves of Fig.\ \ref{fig_swellingcurves}(a)
replotted using the same colors after being rescaled by their minima.}
\label{fig_R1R0}
\end{figure}

Figures \ref{fig_swellingcurves}(a)--(c) show the swelling curves
measured for POPC vesicles of broadly distributed sizes due to the
permeation of three different solutes: urea, glycerol, and ethylene
glycol. Each of these curves has the typical shape shown in Fig.\ 
\ref{fig_swellingcurve}, yet the polydispersity of the vesicles
flattens them once they are displayed together. This is demonstrated
in Fig.\ \ref{fig_R1R0}, where we have replotted the swelling curves
for urea after rescaling the ordinate of each of them by its minimum
radius, $R_0$. Figure \ref{fig_R1R0} shows that such a simple size
rescaling does not collapse the data onto a master curve.  A more
detailed analysis of the temporal swelling curves requires a theory,
which is presented in the next section. In Fig.\ 
\ref{fig_swellingcurves} it is apparent, nonetheless, that there is no
correlation between the transition width (sharpness of the curve's
minimum) and the vesicle size. We return to this point in Sec.\ 
\ref{sec_analysis}.

\begin{figure}[tbh]
\vspace{0.5cm}
\centerline{\resizebox{0.4\textwidth}{!}{\includegraphics{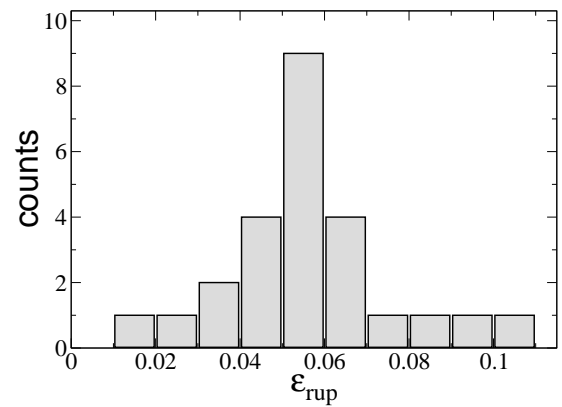}}}
\caption{Histogram of measured rupture strains for POPC vesicles.
  The distribution has a mean of $0.055$ and standard deviation of
  $0.02$.}
\label{fig_rupture}
\end{figure}

A dataset that can be presented without further analysis is the one
for $\erup$, the critical surface strain at rupture, calculated from
the directly measured values of $R_0$ and $R_{\rm rup}$ according to
Eq.\ (\ref{strain}). Since rupture is a result of pore formation,
which is a nucleated event \cite{Litster1975,Farago2005}, the critical
strain as measured in individual vesicles is a stochastic variable.
Figure \ref{fig_rupture} shows the histogram of measured rupture
strains. From this distribution we get $\erup=0.055\pm0.02$, \ie about
5 percent, which agrees with known values for the maximum sustainable
strain of lipid bilayers \cite{Evans1990}.

\section{Theory}
\label{sec_theory}

Models of vesicles usually focus on the statistical mechanics of the
membrane under the appropriate constraints, the starting point being
the Helfrich Hamiltonian
\cite{Helfrich1984,Milner1987,Dobereiner:1997,Seifert1997,Safran}. In
the current work we choose a different, more general description,
based on the thermodynamics of small systems \cite{thermo}. This is
because the phenomenon of main interest here\,---\,the transition
between the ironing and stretching stages\,---\,arises from
competition between volume and surface effects, which is independent
of the detailed interactions. Since the transition is of a mean-field
type \cite{PRL08}, such a thermodynamic description should be
sufficient.

We consider a dilute solution enclosed in a membrane vesicle.  This
system is in thermal and pressure contact with a much larger solution
having temperature $T$ and pressure $\po$. The system is at
thermodynamic equilibrium. Thus, its temperature is $T$.  However,
since we deal with strong surface effects, the system's equilibrium
pressure may deviate from the outer pressure $\po$. The enclosed
solution contains $Q$ solute molecules.\footnote
{In the experiment described above the vesicle actually contains two
  solutes\,---\,one that permeates the membrane and another that is
  trapped inside the vesicle. Analysis for an ideal solution,
  containing $Q_1$ molecules of the former and $Q_2$ molecules of the
  latter, leads to identical results once we take $Q=Q_1+Q_2$.}
Although $Q$ in the experiment changes with time, we assume, because
the solute has much lower permeability than water, that the system is
at all times in quasi-equilibrium, given the instantaneous number of
enclosed solute molecules, $Q(t)$.  Due to the fast water exchange,
we treat both the volume $V$ and surface area $A$ of the vesicle as
free, independent thermodynamic variables.  Alternatively, one can
define a reduced volume,
\begin{equation}
  v(V,A) = 6\sqrt{\pi}\frac{V}{A^{3/2}} \in (0,1),
\label{v}
\end{equation}
which characterizes the deviation of the vesicle from a perfect
spherical shape, and consider $v$ and $A$ as the thermodynamic
variables.

We divide the Gibbs free energy of the system into three-dimensional
(volume) and two-dimensional (surface) contributions:
\begin{equation}
  G = G_{\rm 3D} + F_{\rm 2D}.
\label{G}
\end{equation}
Taking the enclosed solution to be ideal and dilute, we write its
Gibbs free energy as
\begin{equation}
  G_{\rm 3D} = \kB T Q \left[ \ln\left( \frac{Q\vw}{V} \right)
  - 1 \right] + \po V,
\label{G3D}
\end{equation}
where $\vw$ is the molecular volume of water.

As to the surface, we distinguish between in-plane (stretching) and
out-of-plane (bending) degrees of freedom, and assume that the vesicle
is sufficiently large such that the two are decoupled, \ie contribute
additively to the Helmholtz free energy,
\begin{equation}
  F_{\rm 2D} = F_{\rm s} + F_{\rm b}.
\end{equation}
We further assume that the membrane's in-plane free energy is minimum
for a certain relaxed area, $A_0$, and expand it to leading order
about that minimum,
\begin{equation}
  F_{\rm s} = \frac{K}{2A_0} (A-A_0)^2,
\end{equation}
where $K$ is the membrane's stretching modulus. In the bending free
energy we include only the contribution from undulation entropy,
which, assuming small fluctuations about a spherical shape, is given
by \cite{PRL08}
\begin{equation}
  F_{\rm b} = -\frac{1}{2} N \kB T \ln(1-v),
\label{bendentropy}
\end{equation}
where $N$ is the total number of independent bending modes
contributing to the membrane's thermodynamics.  Equation
(\ref{bendentropy}), which describes the suppression of bending
fluctuations as the vesicle approaches the maximum volume-to-area
ratio ($v\rightarrow 1$), is the only statistical-mechanical input to
the theory. It is analogous to an equipartition principle; here each
bending mode contributes an equal amount of $(\kB T/2)|\ln(1-v)|$ to
the free energy \cite{PRL08}.  Thus, this expression holds quite
generally for nearly spherical vesicles, regardless of their detailed
properties. Equation (\ref{bendentropy}) also provides the crucial
coupling between $G_{\rm 3D}$ and $F_{\rm 2D}$, since $v$ depends on
both $V$ and $A$ according to Eq.\ (\ref{v}). 

The number of modes affecting the thermodynamics, $N$, which should be
proportional to the number of lipid molecules in the membrane (\ie to
the membrane's relaxed area $A_0$), is {\it a priori} unknown. Further
discussion of this number, its dependence on membrane's parameters,
and its link to other theoretical treatments of membrane fluctuations,
is deferred to Sec.\ \ref{sec_discuss}. Here we prefer to leave it
unspecified and extract it from the experiment. The number of modes is
used to define an intensive area, or effective patch size, $a\equiv
A/N$, having the relaxed value $a_0\equiv A_0/N$.

One could add, of course, various other contributions to the free
energy of the vesicle, such as the membrane's bending energy,
surface--surface and surface--solute interactions (\eg electrostatic
ones), gravitational energy, {\it etc.}. All these, however, do not
significantly change when the vesicle becomes spherical and begins to
stretch. In other words, in the limit (to be studied below) where
$(1-v)$ becomes singular, these contributions are non-singular, and,
hence, cannot affect the critical behavior in the transition region.

Equations (\ref{G})--(\ref{bendentropy}) define the Gibbs free energy
of our finite-size system as a function of $(T,\po,Q,N)$ and also $V$
and $A$. The equilibrium free energy as a function of $(T,\po,Q,N)$
alone is obtained by minimizing $G$ with respect to the volume and
area of the vesicle. Note that by using this procedure we circumvent
altogether the subtle and controversial issue of membrane tension.
(See, \eg Refs.\ \citenum{Farago2004,Fournier2008,Schmid2011,thermo} and
references therein.) The Laplace tension and surface pressure of the
membrane emerge naturally from the minimization with respect to $V$
and $A$, respectively. That is also the reason why we have formulated
the Helmholtz (rather than Gibbs) free energy of the surface, to avoid
specifying any thermodynamic variables pertaining to the surface other
than $T$, $A$, and $N$. Another related subtlety is that
$K=A(\pd^2F_{\rm 2D}/\pd A^2)$ at fixed $v$ (rather than fixed $V$),
otherwise there would be another contribution to the stretching
modulus from changes in the volume-to-area ratio.

The minimization leads to the following two equations:
%
\begin{eqnarray}
  \frac{1}{2}\delta_N \frac{v}{1-v} + \left(\frac{a}{a_0}\right)^{3/2} v
  &=& \frac{Q}{\Qc}
\label{minimization1}\\
  \frac{2\delta_N}{3\delta_K} \frac{a(a-a_0)}{a_0^2} +
  \left(\frac{a}{a_0}\right)^{3/2} v &=& \frac{Q}{\Qc},
\label{minimization2}
\end{eqnarray}
%
from which one can calculate $v$ and $a$. In Eqs.\
(\ref{minimization1}) and (\ref{minimization2}) we have defined
\begin{equation}
  \Qc \equiv \frac{\po V_0}{\kB T},\ \ \
  V_0 \equiv \frac{A_0^{3/2}}{6\sqrt{\pi}},
\label{Qc_def}
\end{equation}
along with two small parameters,
\begin{equation}
  \delta_N \equiv \frac{N}{\Qc} \sim N^{-1/2},\ \ \
  \delta_K \equiv \frac{\kB T}{K a_0} \sim K^{-1},
\end{equation}
the first related to the vesicle's finite size, and the other to the
membrane's finite stretchability.

Before proceeding, it is instructive to examine Eqs.\
(\ref{minimization1}) and (\ref{minimization2}) in two limits. The
first is the ordinary thermodynamic limit of infinite system size
($N\rightarrow\infty$, $\delta_N\rightarrow 0$) while keeping the
stretching modulus finite.  In this case Eqs.\ (\ref{minimization1})
and (\ref{minimization2}) become degenerate and yield the expected
equilibration of inner and outer pressures, $\kB TQ/V=\po$. The second
limit is that of an unstretchable membrane ($K\rightarrow\infty$,
$\delta_K\rightarrow 0$). In this case Eq.\ (\ref{minimization2})
imposes $a=a_0$, and Eq.\ (\ref{minimization1}) then yields the
behavior that was studied in Ref.\ \citenum{PRL08}, with criticality
of $v$ as $N\rightarrow\infty$. Outside these two limits we get from
these equations the relation
\begin{equation}
  \delta_K \frac{v}{1-v} = \frac{4a(a-a_0)}{3a_0^2},
\label{v_a_relation}
\end{equation}
which demonstrates the actual interplay between the volume-to-area
ratio and surface strain. We see that the vesicle can never attain a
perfect spherical shape ($v=1$), because this would require unphysical
strain ($a\rightarrow\infty$). The deviation of the area from its
relaxed value is inversely proportional to the deviation of the shape
from a perfect sphere. This relation becomes more sensitive the
smaller the value of $\delta_K$\,---\,\ie the vesicle should be strongly
swollen, $1-v\sim\delta_K\ll 1$, to get an appreciable strain
$(a-a_0)/a_0$.

To present the detailed behavior of the transition, we define a
control parameter $q$, proportional to the number of enclosed solute
molecules, and an order parameter $M$, measuring the deviation from
a spherical shape,
\begin{equation}
  q \equiv Q/\Qc -1,\ \ \ M\equiv 1-v.
\label{q_def}
\end{equation}
Solving Eqs.\ (\ref{minimization1}) and (\ref{minimization2}) in the
vicinity of the transition, we get
\begin{eqnarray}
  &&M(q) = \Delta \tM\left(q/\Delta\right),\ \ \
  \Delta = \left[ 2\delta_N + (9/2)\delta_K \right]^{1/2}
 \nonumber\\
  &&\tM(x) = \left(\sqrt{1+x^2} - x \right)/2.
\label{scaling1}
\end{eqnarray}
Thus, as $q$ increases from negative to positive values, $M$ crosses
over from appreciable values, $M\simeq|q|$, to very small ones,
$M\simeq\Delta/(4q)$. The crossover occurs over a range of $q$ defined
by $\Delta$. In the limit $\Delta\rightarrow 0$ the point $(q=0,M=0)$
becomes a singular corner where $dM/dq$ undergoes a discontinuous
jump. The finite size of the vesicle and the stretchability of the
membrane both contribute to the broadening of the transition. If we
take $\delta_K=0$, Eq.\ (\ref{scaling1}) properly coincides with the
critical behavior in the case of an unstretchable membrane
\cite{PRL08}. Which of the two broadening effects dominates in
practice is determined by the ratio
\[
  \frac{\delta_N}{\delta_K} \sim \frac{K/R_0}{\po},
\]
where the right-hand side reflects the competition between surface (numerator)
and volume (denominator) effects.

Substituting Eq.\ (\ref{scaling1}) in Eq.\ (\ref{v_a_relation}), we
obtain for the surface strain,
$\epsilon\equiv(a-a_0)/a_0=(3/4)\delta_K M^{-1}$. Yet, since the
scaling function satisfies $[\tM(x)]^{-1}=4\tM(-x)$, we can rewrite
this result as
\begin{equation}
  \epsilon(q) = \frac{2}{3[1+(4/9)(\delta_N/\delta_K)]}
  \Delta \tM(-q/\Delta).
\label{scaling_epsilon}
\end{equation}
Thus, up to a prefactor, the strain's behavior is a mirror image of
that of $M$. As $q$ increases from negative to positive values,
$\epsilon$ changes from very small values, $\sim\Delta/|q|$, to
appreciable ones, $\sim q$. Similarly, combining Eqs.\
(\ref{minimization1}), (\ref{v_a_relation}), and (\ref{scaling1}), we
get for the Laplace tension,
\begin{equation}
  \gamma(q) = \frac{R}{2}\left(\frac{\kB TQ}{V}-\po\right)
  \simeq \frac{3\kB T}{4a_0}[M(q)]^{-1} = K\epsilon(q),
\label{gamma}
\end{equation}
\ie the surface tension follows the same behavior as the strain,
becoming appreciable only for sufficiently large, positive $q$.
Equation (\ref{gamma}) implies that within the transition region the
tension is in the so-called linear regime
\cite{Helfrich1984,Evans1990}, where it is proportional to the surface
strain. Substituting the rupture strain found above,
$\erup\simeq0.05$, and $K\simeq240$ mN/m \cite{Marsh2006}, we obtain a
rupture tension of order $10$ mN/m, which is in line with previously
reported values of $5$--$10$ mN/m \cite{Evans1990}.

Equation (\ref{scaling1}) [equivalently, (\ref{scaling_epsilon}) or
(\ref{gamma})] describes a law of corresponding states for osmotically
swollen, nearly spherical vesicles. It predicts that, upon rescaling,
the swelling (\ie volume-to-area ratio, strain, or tension) of such
vesicles, regardless of their size, composition, and the nature or
concentration of solute, could be collapsed onto a single universal
curve defined by the function $\tM(x)$. Thus, despite the many
physical parameters affecting the process, only two parameters
actually suffice to completely characterize the osmotic swelling of
any nearly spherical vesicle\,---\,the location of the transition,
$\Qc$, and its width, $\Delta$.

An interesting consequence of the competition between surface and
volume effects is found when we examine the vesicle's swelling beyond
the transition, well into the stretching stage. In this case,
$q\gg\Delta$, Eq.\ (\ref{scaling_epsilon}) becomes
\begin{equation}
  \epsilon = \frac{2}{3} \frac{q}{1+(4/9)\delta_N/\delta_K} = 
  \frac{2}{3} \frac{\kB T Q/V_0 - \po}{\po + (4/3)K/R_0}.
\label{epsilon_stretch}
\end{equation}
Upon replacing the surface strain of a spherical vesicle with its
relative volume change, $(V-V_0)/V_0=(3/2)\epsilon$, and the outer
pressure of a dilute solution with its concentration, $\po=\kB T\co$,
Eq.\ (\ref{epsilon_stretch}) reproduces a known result
\cite{Mally2002}, which was used in earlier analyses of the stretching
stage \cite{Mally2002,CSB08}. As we have seen above, if
$\delta_N/\delta_K\ll 1$, the finite stretching modulus determines the
transition width $\Delta$ [Eq.\ (\ref{scaling1})]. At the same time,
according to Eq.\ (\ref{epsilon_stretch}), the relatively small value
of $K$ in this case makes the swelling beyond the transition
insensitive to $K$. Being governed by volume effects, the growth in
the stretching stage is then simply proportional to the increasing
number of enclosed molecules \cite{CSB08}.

\section{Data analysis}
\label{sec_analysis}

Let us first check which of the broadening factors\,---\,finite size
or stretchability\,---\,is dominant in the experiment. The stretching
modulus of a POPC bilayer is $K\simeq 240$ mN/m \cite{Marsh2006}. Our
vesicles have $R_0\sim 20$--$50$ $\mu$m, and the outer concentration
is $\co\sim 0.1$--$0.2$ M.  We get $\delta_N/\delta_K\sim K/(R_0\kB T
\co)\sim 0.01$--$0.1$. Hence, the transition width is governed by the
finite stretching modulus and is expected, therefore, to be
independent of vesicle size,
\begin{equation}
  \Delta \simeq \left(\frac{9}{2}\delta_K\right)^{1/2}
  = \left(\frac{9\kB T}{2Ka_0}\right)^{1/2}.
\label{Delta}
\end{equation}
As noted above, this implies also that the stretching stage following
the transition is insensitive to the value of $K$, with 
$\epsilon(q\gg\Delta)\simeq (2/3)q$.

To further check the theoretical predictions of the preceding section,
we need to relate them to the experimental observables demonstrated in
Fig.\ \ref{fig_swellingcurve}\,---\,\ie the temporal change in the
principal radius of an oblate spheroidal vesicle, $R_1(t)$.

We begin with the time axis. The membrane permeability coefficients of
the examined solutes are of order $P\sim 10^{-2}$ $\mu$m/s
\cite{Paula1996,DordasJMB00,CSB08}, the vesicles are of radii $R_0\sim
20$--$50$ $\mu$m, and the swelling process lasts about $t\sim
10$--$100$ s. We have $(P/R_0)t\sim 0.001$--$0.05$, implying that the
concentration of permeating solute inside the vesicle never exceeds a
few percent of its outer concentration. Hence, throughout the observed
swelling we may assume $dQ/dt=PA_0\co$, leading to
\begin{equation}
  q = (3P/R_0)t + \mbox{const}.
\label{q_t}
\end{equation}
Thus, the time axis, up to a linear transformation that depends on
vesicle size, is equivalent to our control parameter.

Next, we should adapt the theoretical scaling relations, Eqs.\
(\ref{scaling1}) and (\ref{scaling_epsilon}), so that they could be
applied to the measured swelling curves, $R_1(t)$. This calculation is
presented in the Appendix. Given the experimentally measured
dependence, $[R_1(t)/R_0]^3-1\equiv\gexp(t)$, we find that the following
scale and shift transformation:
\begin{eqnarray}
  f(x) &=& (15\Delta/8)^{-1/2} \left\{ \gexp[x-(15^{1/3}/4)\Delta^{-1/3}]
  -x\Delta \right.\nonumber\\
  &&\left.+ 3(15^{1/3}/4)\Delta^{2/3} \right\},
\label{collapse}
\end{eqnarray}
where $x=q/\Delta$ and $q=(3P/R_0)t$, should collapse the data onto
the universal function
\begin{equation}
  f(x) = (\sqrt{1+x^2}-x)^{1/2}.
\label{f}
\end{equation}

\begin{figure}[tbh]
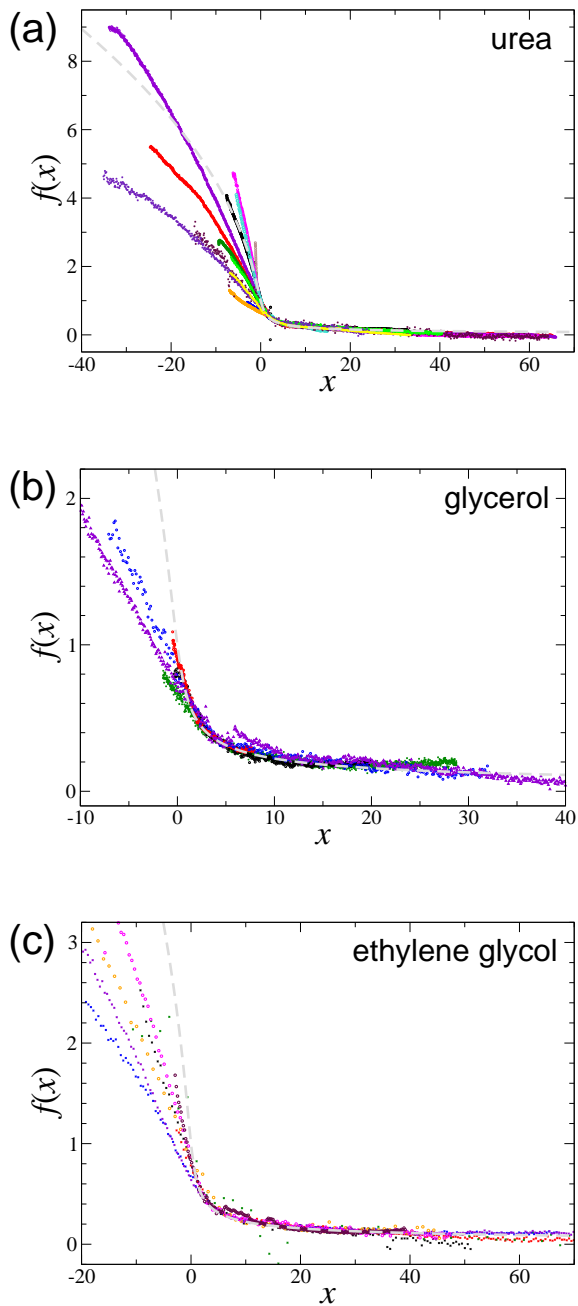

\vspace{0.5cm}
\centerline{\resizebox{0.42\textwidth}{!}{
  \includegraphics{fig5a.eps}}}
\vspace{0.95cm}
\centerline{\resizebox{0.42\textwidth}{!}{
  \includegraphics{fig5b.eps}}}
\vspace{0.95cm}
\centerline{\resizebox{0.42\textwidth}{!}{
  \includegraphics{fig5c.eps}}}
\caption{The experimental data of Fig.\ \ref{fig_swellingcurves}
  replotted using the same colors after being transformed according to
  Eq.\ (\ref{collapse}). The dashed gray line shows the theoretical
  master curve, $f(x)=(\sqrt{1+x^2}-x)^{1/2}$.}
\label{fig_collapse}
\end{figure}

The rescaling scheme defined in Eq.\ (\ref{collapse}), which has been
dictated by the nature of the experiment, is not as elegant as the
theoretical law of corresponding states, Eq.\ (\ref{scaling1}).
Nevertheless, the two are equivalent and similarly straightforward. We
have applied the scheme to the data of Fig.\ \ref{fig_swellingcurves}
while using the permeability $P$ (which merely scales the horizontal
axis) and the transition width $\Delta$ as two fitting parameters. The
results are presented in Fig.\ \ref{fig_collapse}, showing successful
data collapse, within the transition region and above it, onto the
predicted universal curve, Eq.\ (\ref{f}). Theoretically, successful
data collapse is expected in the vicinity of the transition. We obtain
data collapse also above the transition, throughout the stretching
stage, for technical reasons explained in the Appendix. Below the
transition (\ie for sufficiently negative $x=q/\Delta$), the different
swelling curves in Fig.\ \ref{fig_collapse} depart from the universal
behavior. The entropy effects responsible for this departure, which
set in at sufficiently low surface tension, will be described in
Sec.\ \ref{sec_discuss}.

\begin{figure}[tbh]
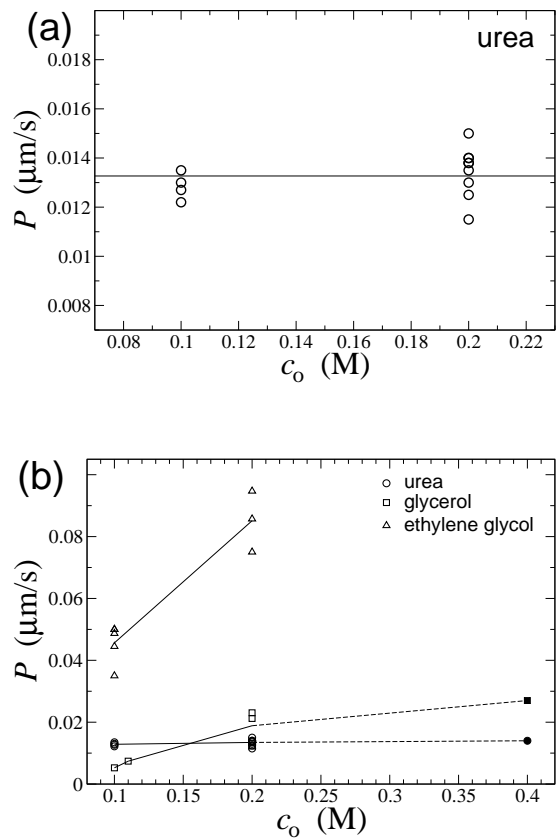

\vspace{0.5cm}
\centerline{\resizebox{0.4\textwidth}{!}{
  \includegraphics{fig6a.eps}}}
\vspace{0.95cm}
\centerline{\resizebox{0.4\textwidth}{!}{
  \includegraphics{fig6b.eps}}}
\caption{Permeability coefficients as a function of outer solute
  concentration for urea (a) and all three solutes (b). Uncertainties
  in individual fitted values are smaller than the scatter for different
  vesicles. Filled symbols at $\co=0.4$~M are results for DOPC
  vesicles taken from Ref.\ \citenum{Paula1996}. The lines are guides
  to the eye.}
\label{fig_permeability}
\end{figure}

In Fig.\ \ref{fig_permeability} we present the fitted values for the
POPC-membrane permeability coefficients of the three examined
solutes. The values for the permeability of urea [Fig.\
\ref{fig_permeability}(a)] are narrowly distributed and
concentration-independent, yielding $P=0.013\pm 0.001$ $\mu$m/s.  This
agrees well with the result of $0.014$ $\mu$m/s, obtained for $0.1$
$\mu$m-radius DOPC vesicles by Paula \etal \cite{Paula1996} using
dynamic light scattering, despite the different phospholipid and the
large difference in the sizes of the studied systems. For glycerol we
get values that significantly increase with glycerol concentration
[Fig.\ \ref{fig_permeability}(b)]: $P=0.0053$, $0.0074$, and $0.019\pm
0.006$ $\mu$m/s for $\co=0.1$, $0.11$, and $0.2$ M, respectively.  The
last value agrees with the value of $0.021\pm 0.008$ $\mu$m/s,
extracted for the same vesicles, at $\co=0.2$ M glycerol, from a
different analysis of the stretch--burst cycles that follow vesicle
rupture \cite{CSB08}.  A glycerol permeability of $0.027$ $\mu$m/s was
reported for $0.1$ $\mu$m-radius DOPC vesicles by Paula \etal
\cite{Paula1996} and by Dordas and Brown \cite{DordasJMB00}, both at
$\co=0.4$~M. These results are not inconsistent with ours, especially
given the concentration dependence.  For ethylene glycol we obtain
more broadly distributed values that depend on concentration [Fig.\
\ref{fig_permeability}(b)]: $P=0.046\pm 0.006$ and $0.085\pm 0.01$
$\mu$m/s for $\co=0.1$ and $0.2$ M, respectively. The only previously
measured permeability for ethylene glycol that we are aware of is
$0.88$ $\mu$m/s, obtained by Orbach and Finkelstein \cite{Orbach:1980}
using conductivity measurements for a flat egg-PC membrane.  This
value is larger than ours by an order of magnitude. We note, though,
the very different conditions under which the two experiments were
conducted. All permeability values reported here, including the
concentration-dependent ones for glycerol and ethylene glycol, are
consistent with those found from the alternative analysis of
stretch--burst cycles (unpublished results).

\begin{figure}[tbh]
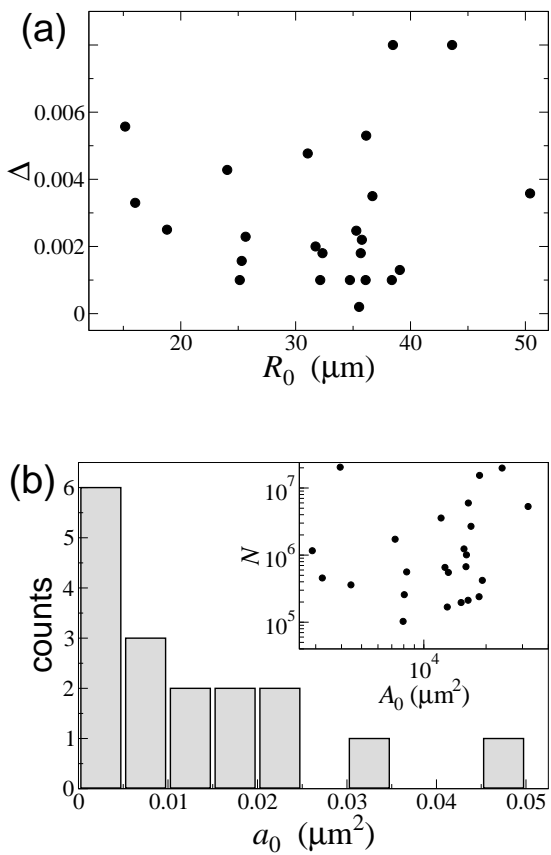

\vspace{0.6cm}
\centerline{\resizebox{0.4\textwidth}{!}{
   \includegraphics{fig7a.eps}}}
\vspace{0.95cm}
\centerline{\resizebox{0.4\textwidth}{!}{
 \includegraphics{fig7b.eps}}}
\caption{(a) Transition width {\it vs}.\ vesicle radius, exhibiting no
  correlation between the two variables. (b) Histogram of effective
  patch size values, obtained from the measured $\Delta$ of panel (a)
  using Eq.\ (\ref{Delta}) and $K=240$ mN/m. Inset shows the resulting
  number of effective bending modes {\it vs}.\ vesicle area.}
\label{fig_Delta}
\end{figure}

We now turn to the second fitting parameter\,---\,the transition
width, $\Delta$. It is found to lie in the range $10^{-3}$--$10^{-2}$
and be uncorrelated with the vesicle size (Fig.\ \ref{fig_Delta}), as
predicted above. The small values of $\Delta$ establish the
self-consistency of our analysis for the crossover between the two
swelling stages as a slightly broadened phase transition, and thus
explain the success of the resulting data collapse. The rather broad
distribution of these values imply that the transition width is
affected by a stochastic variable. Of the three parameters appearing
in Eq.\ (\ref{Delta})\,---\,$T$, $K$, and $a_0$\,---\,only the
effective patch size $a_0$ may be responsible for such stochasticity.
In Fig.\ \ref{fig_Delta}(b) we show the distribution of patch sizes,
as arising from Eq.\ (\ref{Delta}) using the measured $\Delta$ and
$K=240$ mN/m.  The mean patch size is $a_0\simeq 0.01$ $\mu$m$^2$,
which is much smaller than the vesicle area, but much larger than the
molecular size. Such a patch contains about $10^4$--$10^5$ lipids. In
the inset of Fig.\ \ref{fig_Delta}(b) we recast the same data in terms
of the number of effective bending modes, $N=A_0/a_0$. The number of
modes is broadly distributed in the range $10^5$--$10^7$, making its
increasing trend with vesicle area hardly discernible.

\section{Discussion}
\label{sec_discuss}

The experimental results presented above confirm the critical nature
of the osmotic swelling of vesicles, as was first suggested in Ref.\
\citenum{PRL08}. It is important to note that a similar critical
scaling is absent in the case of hydrostatic swelling, as in vesicle
aspiration experiments \cite{Evans1990}. To adapt our formulation to
such a case we should merely replace the solute contribution to the
Gibbs free energy (the first term in Eq.\ (\ref{G3D})), with $(-p_{\rm
in}V)$, where $p_{\rm in}$ is the inner hydrostatic
pressure. Repeating the minimization procedure presented in Sec.\
\ref{sec_theory}, we find that Eq.\ (\ref{v_a_relation}) remains
unchanged, while the swelling order parameter becomes
\begin{equation}
  M = 1-v = \frac{\delta_N}{2(p_{\rm in}/\po-1)}.
\label{Mpin}
\end{equation}
Thus, our formalism applies in this case only for sufficiently large
inner pressures, $p_{\rm in}/\po > 1+\delta_N/2$ (to ensure a positive
$v$). Similar to $q=Q/\Qc-1$, we may define a swelling control
parameter, $q'=p_{\rm in}/\po - 1$, yet its value is restricted to
$q'>\delta_N/2$. Consequently, Eq.\ (\ref{Mpin}) yields a moderate
decrease of $M$ from appreciable to small values, $M\sim
1/q'$\,---\,\ie a gradual approach to a spherical shape. Another way
to demonstrate the non-criticality of the hydrostatic-pressure case is
to recognize that, once both $\po$ and $p_{\rm in}$ are specified, the
surface tension of a nearly spherical vesicle is simply given by
Laplace's law rather than the more elaborate critical expressions of
Eqs.\ (\ref{scaling1})--(\ref{gamma}).  This underlines the special
thermodynamic behavior of osmotically stressed vesicles, as was
indicated before in a more general statistical-mechanical context
\cite{PRE08}. The key difference between osmotically swollen and
directly pressurized vesicles lies in the fact that the pressure
difference of the former depends on the volume for a given number of
encapsulated solute molecules [cf.\ Eq.\ (\ref{G3D})], whereas the
pressure difference of the latter is a given constraint. This makes
the statistical ensembles of $(T,\po,Q,N)$ and $(T,\po,p_{\rm in},N)$
not equivalent \cite{PRE08}.

We have shown how the data collapse for different vesicle sizes can be
utilized to obtain a sensitive measurement of the membrane
permeability coefficients of various solutes. This ability is
particularly well demonstrated in the case of urea (Fig.\ 
\ref{fig_permeability}(a)). Overall, the permeability values that we
have obtained are in accord with previously published ones. (See the
detailed account in the preceding section and Fig.\ 
\ref{fig_permeability}(b).) In addition, we have found an increasing
concentration dependence of the permeability for glycerol and ethylene
glycol. While we are not aware of an earlier report of such an effect,
it is in line with strong evidence for the affinity of polyols to
lipid headgroups \cite{Krasteva2001,Pocivavsek2011} and may account
for the large range of values reported in the literature for the
permeability coefficients of these solutes. The affinity should make
the solute concentration adjacent to the membrane larger than in the
bulk solution, leading to an apparent permeability coefficient which
is larger than the actual one. More extensive measurements of the
apparent permeability as a function of concentration might allow an
extraction of the solute--membrane affinity parameters. At higher
concentrations the solute may also disrupt the membrane structure,
thus affecting the actual permeability \cite{Patel2010} and possibly
other membrane properties \cite{Vitkova2006}.

The transition width, and the corresponding effective membrane patch
$a_0$, have been found to be broadly distributed (Fig.\
\ref{fig_Delta}). The origin of this stochasticity is not yet
clear. It may be that $a_0$ is sensitive to the presence of membrane
impurities. Nevertheless, the analysis of transition widths has
enabled us to obtain information regarding the number of independent
bending modes, $N$, which contribute to the thermodynamics of the
vesicle in the transition region. To our knowledge this variable has
not been experimentally accessed before. For the vesicles used here,
having $R_0\simeq 20$--$50$ $\mu$m, we have found numbers $N\sim
10^5$--$10^7$, which are large but much smaller than the number of
lipid molecules in the vesicle (about $10^{10}$). 

To account for these intermediate values of $N$ we should link our
thermodynamic formulation to the more detailed statistical mechanics
of fluctuating vesicles. The definition of $N$ in Sec.\ 
\ref{sec_theory} has been through the surface entropy of a nearly
spherical vesicle, Eq.\ (\ref{bendentropy}), where each of the $N$
modes contributes an equal amount of $(\kB T/2)|\ln(1-v)|$ to the free
energy \cite{PRL08}.  Examining more closely the partition function of
a fluctuating vesicle within Helfrich's model \cite{Milner1987}, we
find that these modes, whose contribution is singular in $(1-v)$,
belong to the small-wavenumber (tension-dominated) portion of the
fluctuation spectrum. Their number is
\begin{equation}
  N \simeq \gamma R_0^2/ \kappa,
\label{N}
\end{equation}
where $\kappa$ is the membrane's bending rigidity. Equations
(\ref{scaling_epsilon}), (\ref{gamma}), and (\ref{N}) then allow a
theoretical estimate of $N$. At the transition we have $\gamma\simeq
(2/3)K\Delta\tM(-1)$ (cf.\ Fig.\ \ref{fig_collapse}), which for
$K\simeq 240$ mN/m \cite{Marsh2006} and $\Delta\simeq 0.004$ (Fig.\ 
\ref{fig_Delta}(a)) yields $\gamma\simeq 0.8$ mN/m. This value of
about 1 mN/m is in line with that obtained for nearly spherical
vesicles in aspiration experiments \cite{Evans1990}. Substituting it,
together with $\kappa\simeq 10\kB T$ \cite{Marsh2006} and $R_0\simeq
20$--$50$ $\mu$m, in Eq.\ (\ref{N}), we get $N\sim 10^7$, which is in
the range of values that we have found experimentally.

Equation (\ref{N}) implies, in fact, that $N$ is not constant but
increases with the degree of swelling (\ie with $\gamma$). Hence,
sufficiently far below the transition $\gamma$ will be too small, and
the theory should fail.  (Far beyond the transition, on the other
hand, the behavior is dominated by surface strain and unaffected by
these entropy considerations.) Combining Eqs.\ (\ref{gamma}) and
(\ref{N}), we find that the theory should break down for $M\gtrsim
3\kB T/(16\pi\kappa)\simeq 0.006$. With $\Delta\simeq 0.004$ this
occurs for $x\lesssim -1.3$, which is consistent with the region where
the data collapse begins to fail (see Fig.\ \ref{fig_collapse}).


Thus, the assumption of a swelling-independent $N$ is the main
limitation of the model and should be improved in future studies.
Other assumptions are less significant. For example, the relaxed area
of the vesicle, $A_0$, is actually slightly smaller than the assumed
value of $4\pi R_{\rm min}^2$, where $R_{\rm min}$ is the minimum of
the experimental $R_1(t)$ curve.  The calculation given in the
Appendix implies that $R_0/R_{\rm min}\simeq
1-(15^{1/3}/4)\Delta^{2/3}$. For our broadest transitions this
deviation amounts to less than 3\%. As shown in Ref.\ \citenum{PRL08},
accounting for additional factors such as a non-ideal solute does not
affect the universal behavior in the vicinity of the transition.

\subsection*{Acknowledgments}

We thank Oded Farago, Rony Granek, and Luka Pocivavsek for helpful
comments. EH and HD acknowledge the Donors of the American Chemical
Society Petroleum Research Fund for support of this research (Grant
No.\ 46748-AC6).

\section*{Appendix}
\setcounter{equation}{0}
\renewcommand{\theequation}{A\arabic{equation}}

In this appendix we adapt the theoretical scaling relation obtained in
Sec.\ \ref{sec_theory} to the observables in the experimental swelling
curves.  The result is the scaling transformation that should be
applied to the experimental data to achieve data collapse.

Consider an oblate spheroid of principal radii $R_1$ and $R_2<R_1$,
and small eccentricity $e=[1-(R_2/R_1)^2]^{1/2}\ll 1$.  The volume of
the spheroid is $V=(4\pi/3)R_1^2R_2$, and its area is
$A=\pi\{2R_1^2+(R_2^2/e)\ln[(1+e)/(1-e)]\}$. From this we get $v\simeq
1-e^4/15+O(e^6)$, leading to $e^2\simeq(15M)^{1/2}+O(M)$, and
$R_2/R_1=(1-e^2)^{1/2}\simeq 1-(15M)^{1/2}/2+O(M)$. Now we have, on
the one hand, $V/V_0=(A/A_0)^{3/2}v=(1+\epsilon)^{3/2}v$, and, on the
other hand, $V/V_0=(R_1/R_0)^3(R_2/R_1)$. Equating these two
expressions while using the result above for $R_2/R_1$, we obtain
\begin{equation}
  (R_1/R_0)^3-1 \simeq (15M)^{1/2}/2 + 3\epsilon/2.
\label{scaling2}
\end{equation}
The correction to this expression is $O(M)$.  In the transition region
this correction, as well as $\epsilon$, are negligible ($\sim\Delta$)
compared to the $M^{1/2}$ term ($\sim\Delta^{1/2}$). Above the
transition, at the point where $\epsilon$ becomes comparable to
$M^{1/2}$, we find $\epsilon\sim M^{1/2}\sim\Delta^{2/3}$, whereas
$M\sim\Delta^{4/3}$.  Thus, Eq.\ (\ref{scaling2}) contains the leading
terms within the transition region as well as above it. Using Eqs.\
(\ref{scaling1}), (\ref{scaling_epsilon}), and the property
$\tM(-x)=\tM(x)+x$, we rewrite Eq.\ (\ref{scaling2}) as
\begin{eqnarray}
  &&(R_1/R_0)^3-1 \simeq (15\Delta/8)^{1/2} f(q/\Delta)
  + q \nonumber\\
  && f(x) = (\sqrt{1+x^2}-x)^{1/2},
\label{scaling3}
\end{eqnarray}
where we have omitted another term of order $M$ and also the ratio
$\delta_N/\delta_K$.

Equation (\ref{scaling3}), as required, contains the experimental
observable $R_1/R_0$ rather than our order parameter $M$.  Yet, we
need to take one last step before we can apply it to the experimental
data. The expression in Eq.\ (\ref{scaling3}) has a minimum at
$[q\simeq(15^{1/3}/4)\Delta^{2/3},(R_1/R_0)^3-1\simeq
3(15^{1/3}/4)\Delta^{2/3}]$, which is shifted from the one that we
have defined for the experimental curve, $[q=0,(R_1/R_0)^3-1=0]$ (cf.\
Fig.\ \ref{fig_swellingcurve}). In other words, the actual relaxed
radius is slightly shifted from the minimum of the measured swelling
curve.  To correct for this $\sim\Delta^{2/3}$ error, we ought to
shift the experimental curves accordingly. If
$[R_1(t)/R_0]^3-1=\gexp(t)$ is the measured dependence, as a function
of $q$, then the following scaling and shifting transformation,
\begin{eqnarray}
  f(x)&=&(15\Delta/8)^{-1/2} \left\{ \gexp[x-(15^{1/3}/4)\Delta^{-1/3}]
  -x\Delta \right.\nonumber\\
  && \left. +\ 3(15^{1/3}/4)\Delta^{2/3} \right\} \nonumber\\
  &&x=q/\Delta,\ \ \ q=(3P/R_0)t,
\end{eqnarray}
should collapse the data onto the universal function $f(x)$ defined in
Eq.\ (\ref{scaling3}). This is the transformation used in Sec.\
\ref{sec_analysis}.

\end{document}